# Thermal enhanced NIR-NIR anti-Stokes emission in rare earth doped nanocrystals

*Chao Mi, Jiajia Zhou\*, Fan Wang, and Dayong Jin*

[1]Institute for Biomedical Materials and Devices (IBMD), Faculty of Science, University of Technology Sydney, NSW 2007, Australia

\*e-mail: jiajia.zhou@uts.edu.au

**Nanoparticles with anti-Stokes emissions enable many sensing applications, but their efficiencies are considerably low. The key to enable the process of anti-Stokes emissions is to create *phonons* and assist the excited photons to be pumped from a lower energy state onto a higher one. Increasing the temperature will generate more *phonons*, but it unavoidably quenches the luminescence. Here by quantifying the number of *phonons* being generated from the host crystal and at the surface of $Yb^{3+}/Nd^{3+}$ co-doped nanoparticles, we systematically investigated mechanisms towards the large enhancements of the *phonon-assisted anti-Stokes emissions from 980 nm to 750 nm and 803 nm. Moreover, we provided direct evidence that moisture release from the nanoparticle surface at high temperature was not a main reason. We further demonstrated that the brightness of 10 nm nanoparticles were enhanced by more than two orders of magnitude, standing in stark contrast to the thermal quenching effect.**

Anti-Stokes emission process, where radiative transitions occur at energy levels higher than that of the absorbed excitation photons, has enabled a wealth of optical techniques, such as micro-imaging[1,2], laser cooling[3,4], coherent anti-Stokes Raman spectroscopy (CARS)[5,6], photon energy upconversion[7,8], and optical thermometry[9,10]. In particular, nanoscopic probes with anti-Stokes emission properties, like the multi-photon upconversion nanoparticles, are highly attractive to many emerging applications in optical imaging and nanoscale sensing, as it can efficiently remove the detection noise from excitation scatterings and fluorescence background[11-16].

Comparing with the multi-photon upconversion process, the phonon-assisted linear anti-Stokes system only needs one lower-energy photon to produce one higher-energy photon, but it is hard to achieve high efficiency when the anti-Stokes shift is relatively large. As illustrated in Figure 1, to promote the unusual anti-Stokes emission process and meet the energy gaps, *phonon*, that represents an excited quantum state of vibration within a material's elastic structure, plays an essential role. Conventional mechanisms to promote more anti-Stokes emissions included the excited state absorption of phonons (Figure 1a), Boltzmann distribution (Figure 1b), and phonon-assisted energy transfer process (Figure 1c), which often happened inside the crystal hosts and required the increased temperature to favour more phonon generations[17,18]. But higher temperature often 'kills' the emission intensity, known as thermal quenching effect. Therefore, the solutions are often around the exploration of more efficient host materials for generating heat-favourable phonons and proper management of thermal quenching effect[19]. It becomes more challenging when developing a nanoscopic anti-Stokes optical probe, as the increased amount of quenchers at the relatively large surface becomes more active in decreasing the brightness of the nanoparticle probes at increased temperature[20].

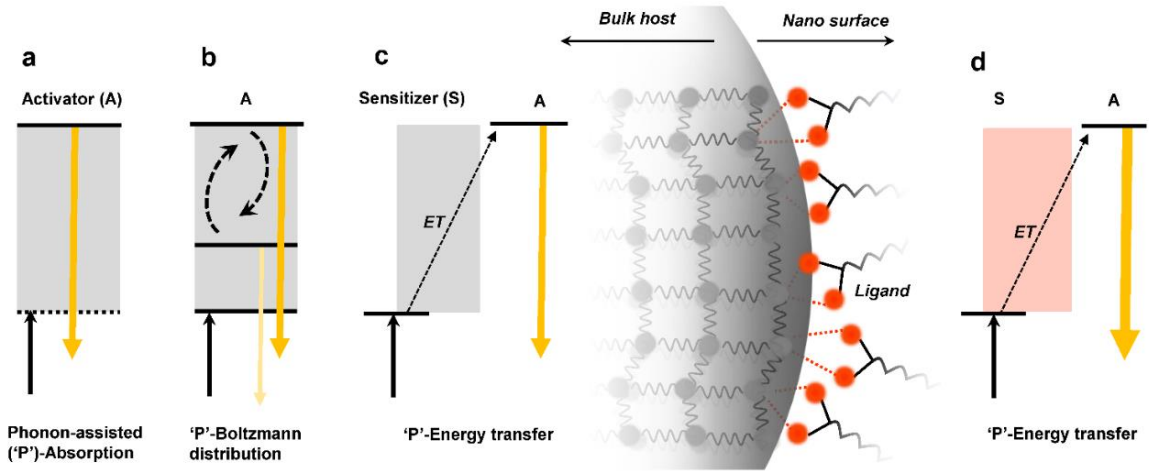

**Figure 1. Phonon-assisted anti-Stokes emission enhancement. a**, Phonon-assisted ('P')-absorption represents the activation of an excited state from a real or imaginary intermediate excited (ground) state by additionally absorbing phonons. This mode enables a type of efficient optical cooler, namely $Yb^{3+}$ doped crystals/glass, to decrease the temperature of the host system. **b**, An emitter has two adjacent excited states, conforming Boltzmann distribution, which allows the phonons to be converted to the higher state with enhanced emission at elevated temperatures. This mechanism enables the design of ratio-metric thermometry. **c**, Phonon-assisted energy transfer process from sensitiser ion to the activator ion. The classical energy transfer couple of $Yb^{3+}$ and $Nd^{3+}$ exhibits a configuration with host phonon accessible energy gap, as most fluorides' and oxides' bonding hold vibration energy of hundreds of wave numbers. **d**, Apart from the bulk host generated phonons, the proposed surface assisted energy transfer process can significantly enhance the one-photon anti-Stokes emission in the nanocrystals.

In the search for high-efficient nanomaterials with large anti-Stokes shift, inorganic nanomaterial systems have been investigated to produce linear anti-Stokes emissions, including the CdSe(Te)/CdS/CdSe quantum dots with an anti-Stokes shift of 110 nm from 680 nm to 570 nm[21], carbon dots from 1100 nm to 984 nm[22], the $CsPbBr_3$ perovskite nanocrystal from 532 nm to 518 nm[23], the CdS nanobelts from 532 nm to 496 nm[24], and h-BN from 610 nm to 565 nm[25]. These material systems commonly suffer from the lack of a pair of suitable excited states as the 'step-stones' to facilitate energy transfer, and thereby their anti-Stokes emissions depend on the contingency of the defects or the bandgap restricted reconstruction.

Lanthanides doped luminescent nanomaterials can be an excellent candidate for linear anti-Stokes emission, due to the characteristic 'ladder-like' multiple excited states of lanthanides, showing inherent long lifetimes in a typical microsecond scale[26]. However, most experimental demonstrations have been limited in bulk materials, e.g. $Yb^{3+}$ and $Nd^{3+}$ co-doped $CaWO_4$, $La_2O_3$, by taking advantage of their maximum phonon energies in the range of ~800 cm$^{-1}$ and ~500 cm$^{-1}$, that can thermally enhance the anti-Stokes emissions of $Nd^{3+}$ by nearly 200-fold at a high temperature of 853 K[27,28]. For nanoparticles, previous research only dabbled in $NaYF_4$ nanocrystals with a thermal enhancement of 1.8-fold at 420 K[29].

Here we report the dramatically enhanced linear anti-Stokes emissions from lanthanide doped luminescent $NaYF_4$:$Yb^{3+}$, $Nd^{3+}$ nanocrystals system. This takes advantage of the phonons generated at the surface of nanomaterials (illustrated in Figure 1d) to break through the limitation in producing the high-efficient linear anti-Stokes emissions.

The schematic energy levels of $Yb^{3+}$ and $Nd^{3+}$ in β-$NaYF_4$ host material (Figure 2a) show an energy mismatches of 1200 cm$^{-1}$ from $Yb^{3+}$: $^2F_{5/2}$ to $Nd^{3+}$: $^4F_{3/2}$ and additional 1100 cm$^{-1}$ to $Nd^{3+}$: $^2H_{9/2}$, $^4F_{5/2}$[30]. To produce the NIR to NIR anti-Stokes emissions from $Nd^{3+}$ with a large shift (~180 nm), several phonons are needed to fill the big energy gaps, if there are only available low-frequency phonons being generated from the host lattice vibration (smaller than 355 cm$^{-1}$)[31]. What is more in this case, at the presence of oleate ligand (OA) bonding with $Yb^{3+}$ at the surface, the more appropriate phonon modes generated at the surface with vibration frequencies within 510-560 cm$^{-1}$ range can benefit the energy transfer[19]. Thus, an enhancement factor of 136-fold was observed from the $^2H_{9/2}$, $^4F_{5/2}\rightarrow {}^4I_{9/2}$ transitions (~803 nm) of 25 nm β-NaYF$_4$: 20% Yb$^{3+}$, 6% Nd$^{3+}$ nanocrystals when the temperature increased from 298 K to 433 K (Figure 2b).

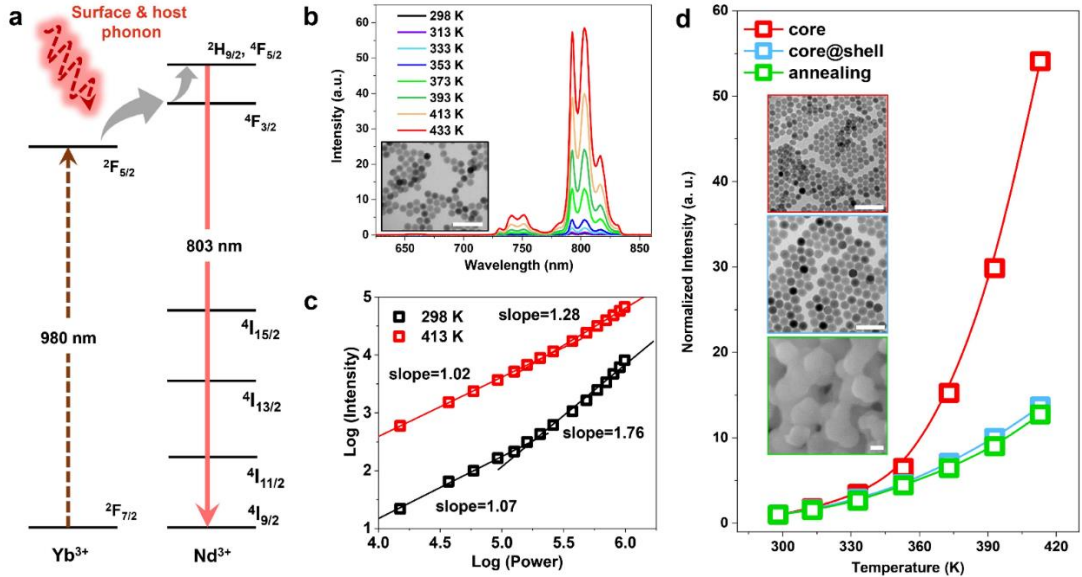

**Figure 2. Phonon-enhanced one-photon anti-Stokes emission in Yb$^{3+}$, Nd$^{3+}$ energy transfer system at elevated temperatures. a**, Schematic energy level diagram showing the surface phonon and host phonon-assisted energy transfer from Yb$^{3+}$ to Nd$^{3+}$. **b**, Intensified anti-Stokes emission spectra of 25 nm β-NaYF$_4$: 20% Yb$^{3+}$, 6% Nd$^{3+}$ nanocrystals with temperature increasing, under the 980 nm continuous wave laser excitation (0.13 W/cm$^2$). Insert TEM image shows the morphology and size of nanocrystals. **c**, Double-log intensity-power slopes of Nd$^{3+}$ 803 nm emission with power density increasing at 298 K and 413 K respectively. **d**, 803 nm emission enhancements of three different samples when the temperature increased from 298 K to 413 K. (980 nm CW excitation 1 W/cm$^2$). Insets show the TEM images of the core and core-shell structure NaYF$_4$: 20% Yb$^{3+}$, 6% Nd$^{3+}$@NaYF$_4$ (core ~20nm, shell ~5 nm) nanocrystals, and SEM image of the annealed NaYF$_4$: 20% Yb$^{3+}$, 6% Nd$^{3+}$ materials. Scale bar: 100 nm.

To exclude what we observed was not the cooperative two-photon upconversion process but the linear one-photon energy transfer[28], we first tested the power dependence of the 803 nm transitions at room temperature, and found as long as the excitation power density was smaller than 2×10$^5$ W/cm$^2$ the anti-Stokes emissions from Nd$^{3+}$ were a linear process (Figure 2c). We then tested the power dependence at 413 K and observed the smaller double-log intensity-power slopes at 413 K than that of 298 K (Figure 2c), which indicated that the phonon-assisted one-photon upconversion was even more pronounced at high temperature because of the availability of more active phonons. As predicted, the cooperative two-photon upconversion process only appeared obviously under an extremely high excitation power at low temperature. These findings have confirmed the linear anti-Stokes emissions from the $Yb^{3+}/Nd^{3+}$ co-doped nanoparticles realized by phonon assistance especially at high temperature.

To demonstrate the important role of surface phonons that makes the anti-Stokes emission more efficient in a thermal field, we also collected the evidence by reverse logic by checking nanomaterials without effective surface ligand. The contrastive samples included the nanocrystals passivated with the inert shell ($NaYF_4$: 20% $Yb^{3+}$, 6% $Nd^{3+}$@$NaYF_4$) to block the phonon assistance from the surface and the annealed samples at 773 K for 3 hours to remove OA. As shown in Figure 2d, we found both control samples without active surface only resulted in ~13-fold intensity enhancement when the temperature increased from 298 K to 413 K, while the core-only 20 nm nanocrystals with active surface exhibited a 55-fold enhancement under the same measurement conditions. Such difference shows the effectiveness of active surface in enhancing the anti-Stokes luminescence.

To further optimise the enhancement factor and achieve strong anti-Stokes emission in $NaYF_4$: $Yb^{3+}$, $Nd^{3+}$ nanoparticles, we prepared a series of ~10 nm nanocrystals to take advantage of their relatively large surface area for surface phonon assistance. TEM results confirmed all the eight samples with various doping concentration had quite similar sizes around 10 nm (Figure 3a and Supplementary Figure 1), and showed a significant amount of enhancement by more than two orders of magnitude after the temperature increased to 453 K (Figure 3a). Interestingly, we found the nanoparticles with higher Yb$^{3+}$ concentration of 60% constantly provided stronger enhancement than the 20% Yb$^{3+}$ doped samples, as the influence caused by different sample sizes can be ignored here, it can only be explained by more bindings between OA and Yb$^{3+}$ ions in high Yb$^{3+}$ doped samples as shown in Figure 1d[19]. Additionally, the quantitative assessment indicated that the enhancement factor had no obvious correlation with the absolute intensity of the nanocrystals (Supplementary Figure 2). Furthermore, by removing OA from both 10 nm and 25 nm samples' surface (confirmed by FTIR in Supplementary Figure 3), we observed the apparent difference that the enhancement in the OA-capped samples was much higher than it in the OA-free samples (Figure 3b).

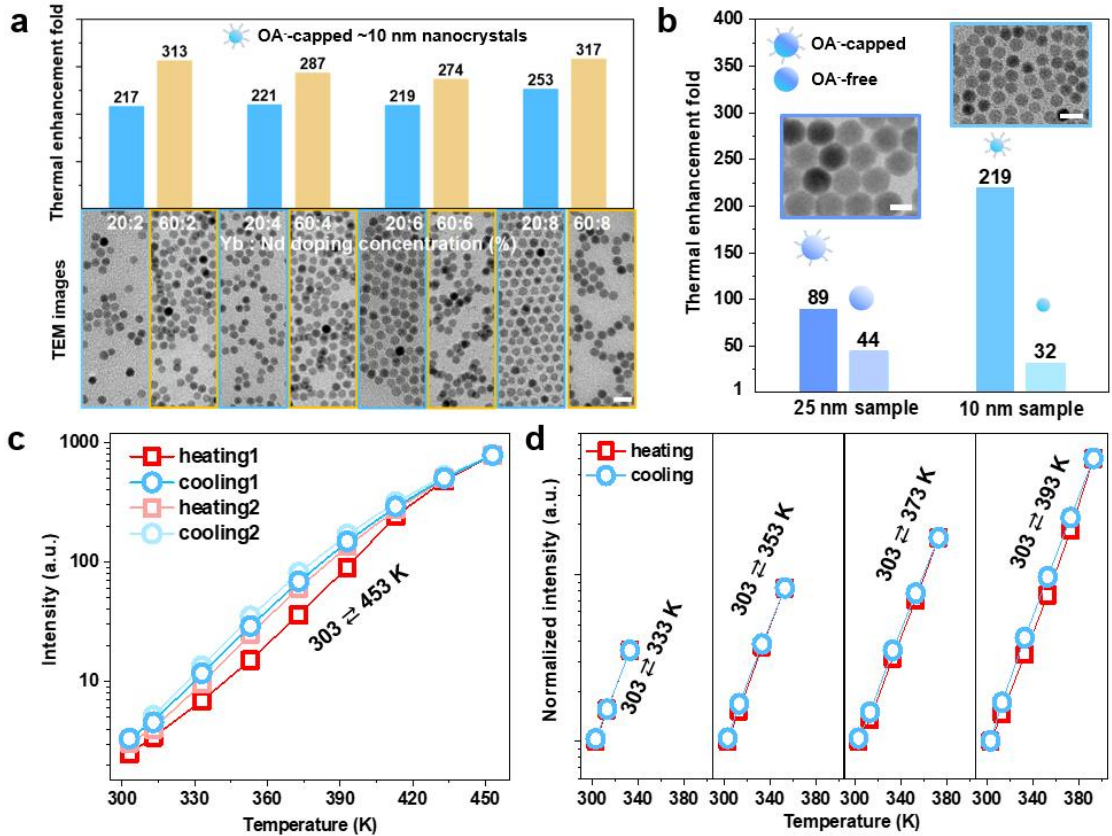

**Figure 3. Thermal enhancement and hysteresis effect in β-NaYF$_4$: Yb$^{3+}$, Nd$^{3+}$ nanocrystals. a**, A series of ~10 nm NaYF$_4$ nanocrystals with varying doping ratios showing more than 200 times enhancement of 803 nm intensity when the temperature increases to 453 K. Scale bar 20 nm. **b**, Intensity enhancement comparisons between OA-capped and OA-removed samples The two sets of samples are 25 nm NaYF$_4$: 20% Yb$^{3+}$, 6% Nd$^{3+}$ and 10 nm NaYF$_4$: 20% Yb$^{3+}$, 6% Nd$^{3+}$ nanocrystals. **c**, Dual-cycle heating-cooling test of the 10 nm NaYF$_4$: 60% Yb$^{3+}$, 8% Nd$^{3+}$ nanocrystals showing the release of hysteresis effect during heating-cooling. **d**, Heating-cooling cycle tests of the 10 nm NaYF$_4$: 60% Yb$^{3+}$, 8% Nd$^{3+}$ nanocrystals with temperatures stopped at different levels before cooling down.

Except the demonstrated surface and host phonon assistance, there are also some other effects may lead to the thermal induced luminescence enhancement[32-34]. To verify whether they coexist for the thermal enhancement, we carried out the heating-cooling cycle test. Figure 3c showed a dual-cycle repeatability test in the air, the reversible emission intensity at room temperature excluded the high temperature annealing effect, which would cause intrinsic structure change of the sample. But the observed hysteresis behaviour did hint the possible activation of the quenched sites, in consideration of the fact that the relatively large surface of nanoparticles could be the parasitic sites for quenchers. As a recent report argued that the thermal induced moisture release may have contributed to the enhancement[32], we have further monitored the hysteresis effect with different maximum temperatures (Figure 3d), and found the hysteresis effect gradually became obvious on higher heating point applied to the sample. These results indicated a possible desorption and reabsorption process of the water molecules during the heating-cooling cycles.

To further prove the phonon assistance from the host and surface could play the dominated role in the anti-Stokes emission enhancement, instead of the moisture release (Figure 4a), we retested the 25 nm and 10 nm samples in a moisture-free environment before and after removing OA on the surface. We evaluated the moisture-free condition by keeping the samples at 453 K and Ar filling environment until achieving stabilized emission intensities (Supplementary Figure

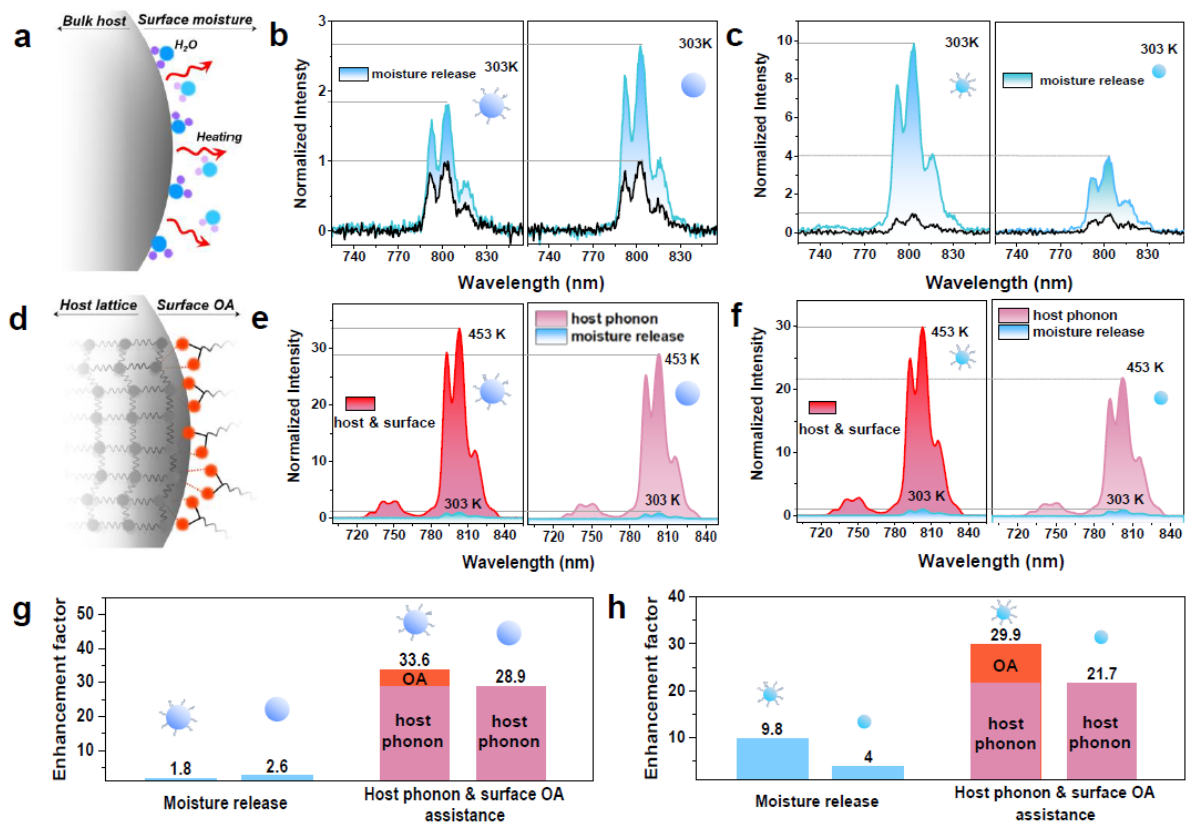

**Figure 4. Quantitative analysis to reveal the roles of host phonon assistance, surface phonon assistance and moisture release in enhancing the anti-Stokes emissions. a**, Schematic of moisture release from the nanoparticle's surface at high temperature. **b** and **c**, Emission spectra of the 25 nm (b) and 10 nm (c) OA-capped and OA-free nanocrystals recorded at 303 K before (black profiles) and after (cyan profiles) moisture release. **d**, Schematic of phonon assistance from both host lattice and active surface. **e** and **f**, Thermal enhancement shown in the emission spectra of 25 nm (e) and 10 nm (f) OA-capped and OA-free nanocrystals from 303 K (cyan profiles) to 453 K (magenta and orange profiles). **g** and **h**, Quantitative analysis of three possible mechanisms that enhance the emissions, including the phonons from the host, the phonons from the surface (OA assistance), and moisture release in the two sets of 25 nm (g) and 10 nm (h) samples. All the tests were carried out in Argon environment with excitation of 0.13 W/cm$^2$ CW 980 nm laser excitation.

4 and 5), and subsequently reduced the temperature to 303 K. As shown in Figure 4b and 4c, we obtained the emission spectra before and after moisture release (Supplementary Figure 6), which corresponded to enhancement factors of 1.8 and 2.6 folds in the 25 nm OA-capped and OA-free samples, and 9.8 and 4 folds for the 10 nm OA-capped and OA-free samples, respectively. Significantly, without the interference from moisture, with additional surface phonon assistance (Figure 4d), we found OA-capped nanoparticles had stronger thermal enhancement than that of OA-free nanoparticles for both 25 nm and 10 nm samples, as shown in Figure 4e and 4f. The quantitative results clearly showed that the assistance of OA contributed to the thermal enhancement from a factor of 28.9 to 33.6 in the 25 nm sample (Figure 4g), and from 21.7 to 29.9 in the 10 nm sample (Figure 4h), which has confirmed the essential role of the surface mode in the enhancement. It was worthy noting that compared to the 25 nm nanoparticles, the 10 nm nanoparticles exhibited stronger effects associated with both mechanisms of the moisture release and OA assistance, which might be due the higher surface to volume ratio of 10 nm sample.

In summary, by achieving the unusual large enhancement for the one-photon anti-Stokes luminescence in $Yb^{3+}/Nd^{3+}$ co-doping nanosystems, at high temperature, and particularly for ultra-small nanoparticles (~10 nm in diameter), this work suggests a large scope for fundamental research on further investigating the phonon-assisted photon conversion process and physical chemistry aspects at the surface of luminescent nanomaterials. This work further suggests new ways in designing anti-thermal-quenching phosphors, high efficient sensors, and temperature-responsive nanophotonics devices for practical applications in the broad areas of bio-imaging, laser cooling, CARS spectroscopy, and nanoscale thermometry.

**Acknowledgement**

The authors acknowledge the financial support from the Australian Research Council Discovery Early Career Researcher Award Scheme (J. Z., DE180100669), National Natural Science Foundation of China (61729501), and Major International (Regional) Joint Research Project of NSFC (51720105015).

# Thermal enhanced NIR-NIR anti-Stokes emission in rare earth doped nanocrystals

Chao Mi, Jiajia Zhou*, Fan Wang and Dayong Jin

*Institute for Biomedical Materials and Devices (IBMD), Faculty of Science, University of Technology Sydney, NSW 2007, Australia*

**Materials and methods**

YCl$_3$·6H$_2$O (99.99%), YbCl$_3$·6H$_2$O (99.99%), NdCl$_3$·6H$_2$O (99.99%), ErCl$_3$·6H$_2$O (99.99%), NH$_4$F (99.99%), NaOH (99.9%), KOH (99.9%), oleic acid (OA, 90%), and 1-octadecene (ODE, 90%) were purchased from Sigma-Aldrich.

**1. Synthesis of NaYF$_4$: Yb$^{3+}$, Nd$^{3+}$ nanocrystals**

NaYF$_4$: Yb$^{3+}$, Nd$^{3+}$ nanocrystals were synthesized by the coprecipitation method. Taken NaYF$_4$: 20% Yb$^{3+}$, 6% Nd$^{3+}$ as an example, 1 mmol RECl$_3$·6H$_2$O (RE = Y, Yb, Nd) with the molar ratio of 74:20:6 were added to a 50 ml flask containing 6 ml OA and 15 ml ODE. The mixture was heated to 180 °C under argon for 30 min to obtain a clear solution and then cooled down to room temperature, followed by the addition of 5 mL methanol solution of NH$_4$F (4 mmol) and NaOH (2.5 mmol). After stirring for 30 min, the temperature was set at 100 °C and the solution was heated under argon for 30 min to remove methanol, and then the solution was further heated to 300 °C for 90 min. Finally, the reaction solution was cooled down to room temperature, and nanoparticles were precipitated by ethanol and washed with cyclohexane, ethanol and methanol for 3 times to get the NaYF$_4$: 20% Yb$^{3+}$, 6% Nd$^{3+}$ nanoparticles.

**2. Controlled synthesis of ~10 nm NaYF$_4$: Yb$^{3+}$, Nd$^{3+}$ nanocrystals**

The family of ~10 nm NaYF$_4$: Yb$^{3+}$, Nd$^{3+}$ nanocrystals with different doping concentration was synthesized by a similar method. Same as above method, the mixture of RECl$_3$·6H$_2$O (RE = Y, Yb, Nd), 6 ml OA and 15 ml ODE was heated to 180 °C under argon for 30 min and then cooled down to room temperature. Then 8 mL methanol solution of 4 mmol NaOH was added to the mixture followed by 15 min stirring. By heating the solution to 100 °C for 15 min, the methanol was removed under argon. Again the mixture was cooled down to room temperature. After that 10 ml methanol solution of 4 mml NH$_4$F was added to the mixture followed by 30 min stirring. Then the temperature was set at 100 °C and the solution was heated for 30 min to remove methanol, next the solution was further heated to 300 °C for 40 min. Finally, the reaction solution was cooled down to room temperature to get the ~10 nm nanocrystals.

**3. Synthesis of ligand-free NaYF$_4$: Yb$^{3+}$, Nd$^{3+}$ nanoparticles.**

The oleate ligand was removed by a simple acid treatment process. Firstly, 4 ml ethanol and 4 ml HCl (0.2M) were added to the as-prepared OA$^-$-capped NaYF$_4$: Yb$^{3+}$, Nd$^{3+}$ precipitate, then the solution was sonicated for 30 min and collected by centrifugation. And the precipitate was further washed for 3 times with 4 ml ethanol and 4 ml deionized water to get the ligand-free nanoparticles.

## 4. Characterization techniques

The morphology of the formed materials was characterized *via* transmission electron microscopy (TEM) imaging (Philips CM10 TEM with Olympus Sis Megaview G2 Digital Camera) with an operating voltage of 100 kV. The samples were prepared by placing a drop of a dilute suspension of nanocrystals onto copper grids.

The morphology of the annealed nanocrystal was characterized via scanning electron microscope (SEM) imaging (Supra 55VP, Zeiss) operated at 20.00 kV.

Fourier transform infrared (FTIR) spectra of the OA$^-$-capped and OA$^-$-free nanocrystals were measured on a FTIR spectrometer (Nicolet 6700, Thermo Scientific).

## 5. Temperature dependent spectra measurement

The temperature dependent spectra were measured with a home-built photoluminescence spectroscopy system. A fiber-coupled 976.5 nm diode laser (BL976-PAG500, controller CLD1015, Thorlabs) with adjustable power up to 500 mW works as the pumping source. By using a temperature heating stage (thermocouple TH100PT, heater HT24S, controller TC200, Thorlabs), the spectra could be measured from room temperature to 453 K by a commercial spectrometer (Shamrock 193i, Andor) with an EMCCD (iXon Ultra 888, Andor) as the detector. For Ar atmosphere measurement, an additional inclosed heating stage (HFS600E-P, Linkam) was used to block air. In addition, the emission signal was filtered by a short pass filter (SPF, FF01-842/SP-25, Semrock) to remove the scattered excitation laser.

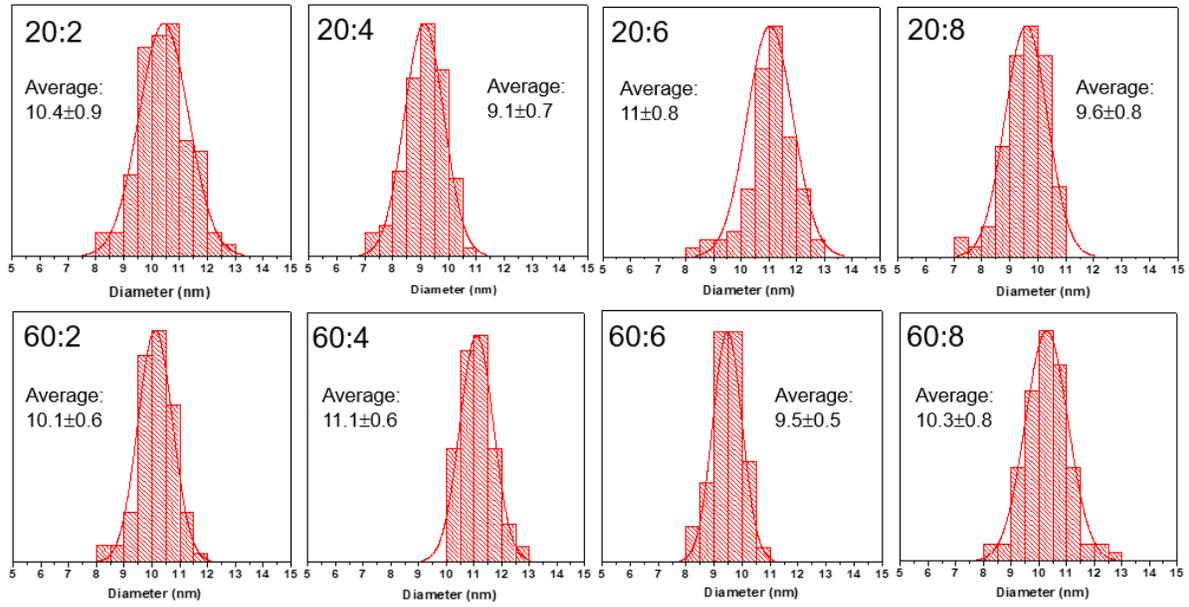

**Supplementary Figure 1. TEM images and size distributions of the ~10 nm NaYF$_4$: Yb$^{3+}$, Nd$^{3+}$ samples.** The doping concentration ratio of Yb$^{3+}$: Nd$^{3+}$ (%) is labeled in each TEM image. Scale bar is 20 nm.

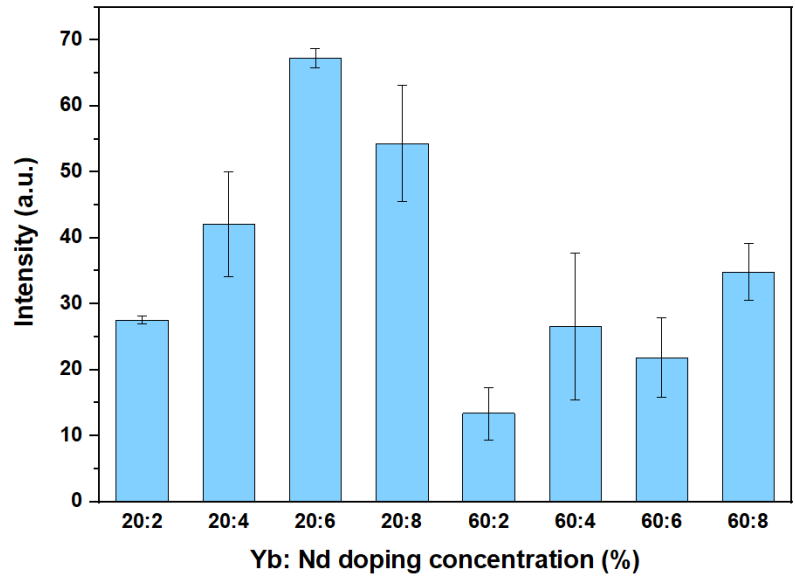

**Supplementary Figure 2. Comparison of 803 nm emission intensity of the ~10 nm NaYF$_4$: Yb$^{3+}$, Nd$^{3+}$ samples.** By controlled synthesis of the different Yb$^{3+}$, Nd$^{3+}$ doping samples with similar size, the emission intensities of the ~10 nm samples are comparable.

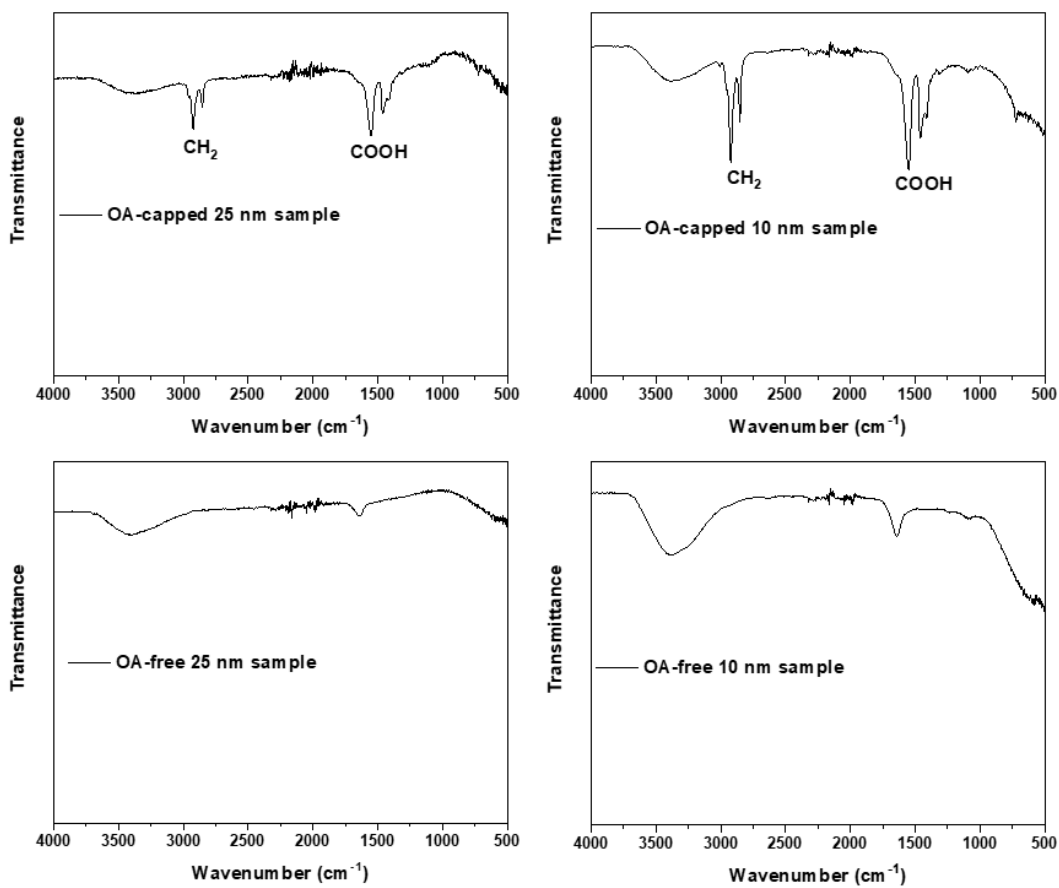

**Supplementary Figure 3. FTIR results of the as-prepared OA-capped and OA-free samples.**
After acid treatment, both $CH_2$ stretching and COOH stretching disappeared, which means surface OA has been removed from 25 nm $NaYF_4$: 20% $Yb^{3+}$, 6% $Nd^{3+}$ and 10 nm $NaYF_4$: 20% $Yb^{3+}$, 6% $Nd^{3+}$ nanoparticles.

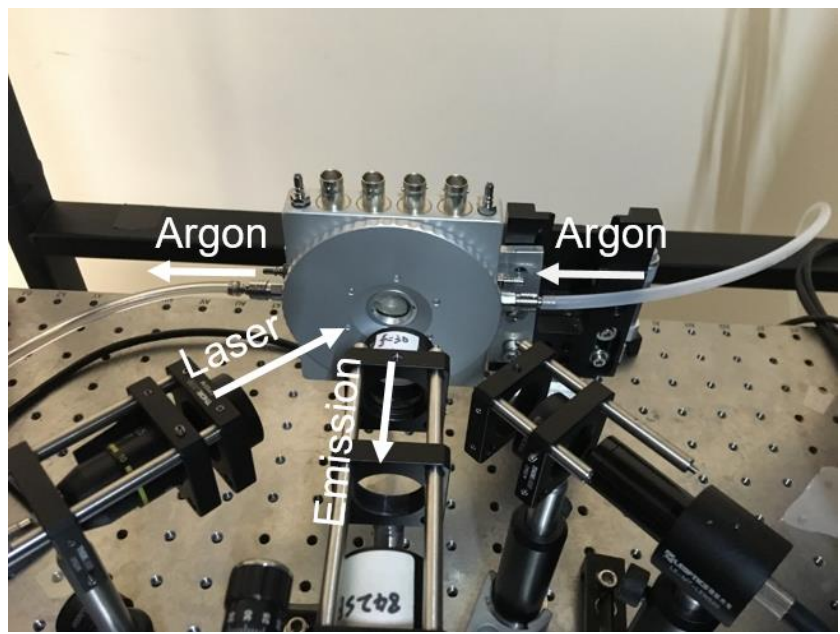

**Supplementary Figure 4. Test system used for Ar atmosphere measurement.** By using an inclosed heating stage, the sample could be measured under Ar to avoid the influence caused by moisture in air.

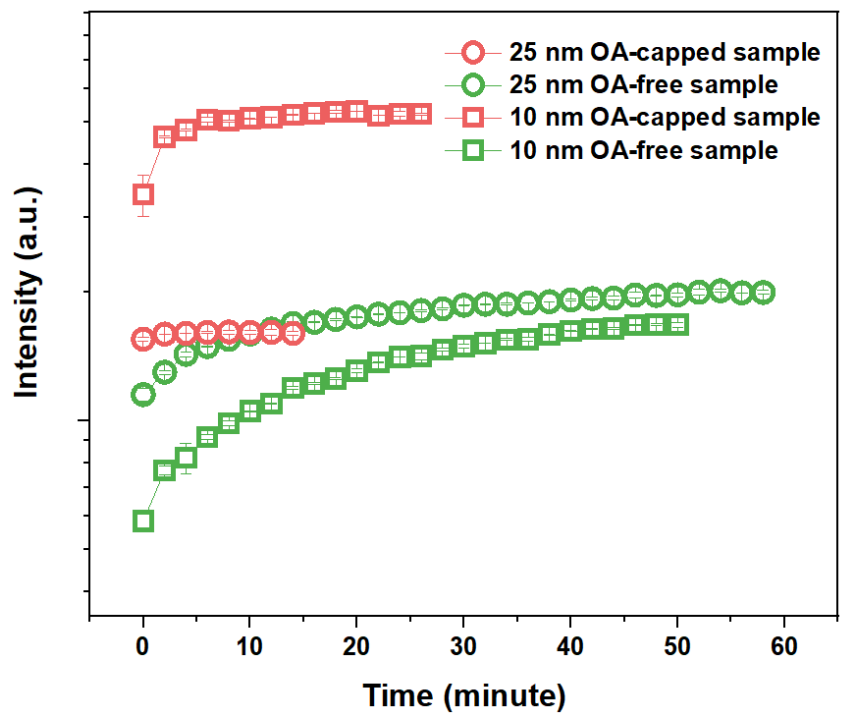

**Supplementary Figure 5. Moisture release process at 453 K.** Time starts from the moment when the set temperature arrives at 453 K, the moisture has been totally removed from the samples until the intensity is stable.

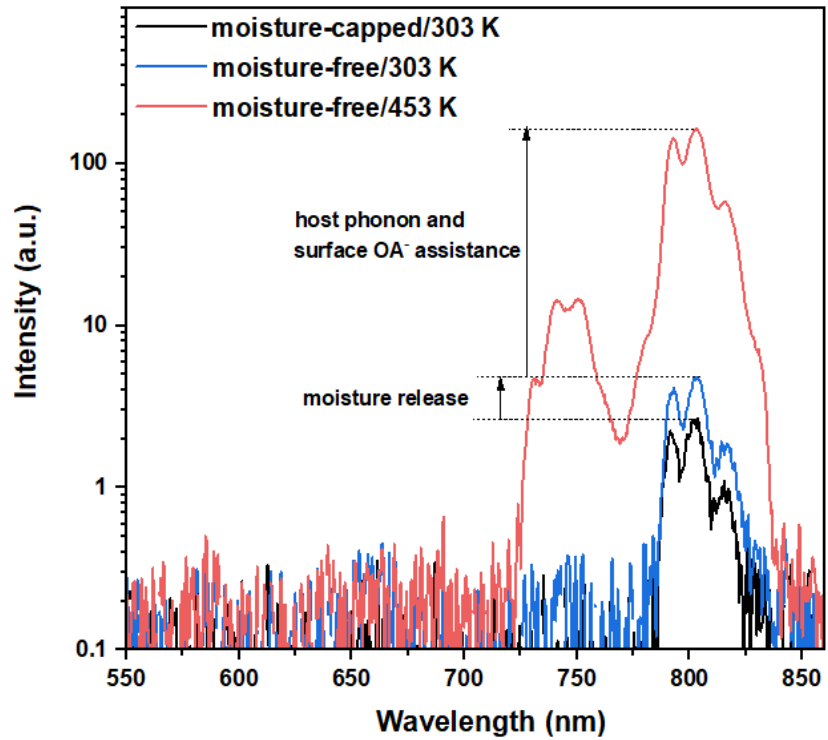

**Supplementary Figure 6. Enhancement factors calculation.** Take the 25 nm OA-capped NaYF$_4$: Yb$^{3+}$, Nd$^{3+}$ sample in Figure 3e for example, the sample exhibits a higher intensity at 303 K after removing the moisture, thus the factor of moisture release equals to the intensity ratio of moisture-capped sample to moisture-free sample at 303 K. Due to the host phonon and surface OA assistance, the moisture-free sample realizes a dramatic emission enhancement at 453 K, thus the factor of host phonon and surface OA assistance is the ratio of the intensity at 453 K to it at 303 K, then by subtracting the factor of host phonon assistance for the 25 nm OA-free sample, the factor of surface OA assistance is also quantified.